\documentclass[12pt]{article}
\usepackage{amsmath,amssymb,mathrsfs,url,graphicx, color}
\usepackage{cite}
\usepackage{comment}

\title{Elimination of cosmological singularities in quantum cosmology by suitable operator orderings}

\newcommand{\dd}{\mbox{d}}

\def\ka{\kappa}

\def\pa{\partial}

\def\al{\alpha}

\def\ka{\kappa}
\def\ii{\textrm i}
\def\ee{\textrm e}

\def\ud{\textrm{d}}

\newcommand{\be}{\begin{equation}}
\newcommand{\en}{\end{equation}}
\newcommand{\bi}{\begin{itemize}}
\newcommand{\ei}{\end{itemize}}

\begin{document}

\small{
\author{Thibaut Demaerel\footnote{{Instituut voor Theoretische Fysica, KU Leuven}} and Ward Struyve$^{*}$\footnote{Centrum voor Logica en Analytische Wijsbegeerte, KU Leuven}}
\date{}
\maketitle

\begin{abstract}
Canonical quantization of general relativity does not yield a unique quantum theory for gravity. This is in part due to operator ordering ambiguities. In this paper, we investigate the role of different operator orderings on the question of whether a big bang or big crunch singularity occurs. We do this in the context of the minisuperspace model of a Friedmann-Lema\^itre-Robertson-Walker universe with Brown-Kucha{\v r} dust. We find that for a certain class of operator orderings such a singularity is eliminated without having to impose boundary conditions.
\end{abstract}

\section{Introduction}
Probably the most conservative approach to quantum gravity is canonical quantum gravity, which arises from applying the usual quantization techniques to general relativity. In canonical quantum gravity, the state is given by a wave functional $\Psi(^{(3)}g,\phi)$ on 3-metrics $^{(3)}g$ and some matter degrees of freedom, say a scalar field $\phi$. The wave equation is not dynamical, since time is absent from it, but amounts to ``constraints" on the wave function. These constraints are the diffeomorphism constraint and the Hamiltonian constraint $\widehat{{\mathcal H}}\psi=0$, also called the Wheeler-DeWitt equation. Due to operator ordering ambiguities in the quantization process, many different forms of these constraints can be obtained, resulting in different quantum theories for gravity. A choice of operator ordering that is often made (mostly in the context of minisuperspace models) is to take the Laplace-Beltrami operator with respect to the DeWitt metric on superspace \cite{christodoulakis86,halliwell91b}. However, other choices remain open, especially in the absence of experimental guidance. The choice of operator ordering may have effect on the physical content of the theory \cite{kontoleon99,steigl06,haga17}. In this paper, we investigate the effect on the possible presence of cosmological singularities.

According to general relativity, singularities like a big bang or big crunch singularity are generically unavoidable. This is the content of the Penrose-Hawking singularity theorems. This is usually taken as a signal that the classical theory breaks down and it is hoped for that a quantum theory for gravity will eliminate such singularities. This has been investigated in simplified models, called minisuperspace models, which arise from applying the usual quantization techniques to symmetry-reduced general relativity by assuming homogeneity and isotropy. However, there is the question of what exactly should be meant by space-time singularities in the context of quantum gravity. In general relativity, singularities arise when geodesics cannot be smoothly extended. In quantum gravity, this notion no longer makes sense, since there is no actual metric field. There is only the wave functional on 3-metrics. Different criteria have been proposed for a singularity. Examples include: that the metric has support on singular 3-metrics, that it is peaked on singular 3-metrics, that the expectation of the scale factor vanishes (see, e.g., \cite{dewitt67a,lund73,gotay83,bojowald01,kiefer04,ashtekar06b,ashtekar06c,ashtekar08,kiefer10}). Even though these criteria may shed some light on the issue of singularities, none of these seems completely satisfactory. For example, demanding that the wave function be zero on singular metrics is often viewed as sufficient for singularity avoidance \cite{dewitt67a,kiefer10}. However, the set of singular metrics is often of measure zero so the amplitude of the wave function at those configurations seems of no relevance. Alternatively, it should perhaps be specified at which rate the wave function vanishes near singularities \cite{blyth75}.  For some of these criteria, the question whether or not the singularities are eliminated also hinges on how the problem of time is dealt with (see, e.g., \cite{gotay83}). 

In this paper, we consider a different criterion, namely that space-time singularities do not occur if and only if there is no quantum (probability) flux into singular metrics. To appreciate this criterion and to contrast it with the alternative prevailing viewpoints, it is useful to consider the following example: In classical mechanics, the Coulomb potential $-Ze^2/r$ is singular at $r=0$. Classically a particle can reach this singularity and its motion cannot be continued thereafter. Is this singularity eliminated in quantum mechanics? In the case of the Dirac equation with the Coulomb potential, the ground state is given by 
\be
\psi_0(r) \sim (2mZ \al r)^{\sqrt{1-Z^2 \al^2} - 1} \ee^{-mZ\al r} . 
\en
Can the quantum particle fall into the singularity of the Coulomb potential? There is no actual point-particle like in classical mechanics, just the wave function. The wave function is nonzero at the origin. It even diverges. So according to some of the proposed notions of a singularity, the singularity is not eliminated. However, quantum mechanically, according to the Born rule, the wave function merely determines a probability amplitude to find the particle somewhere upon detection. The probability (with respect to the measure $\psi^\dagger \psi d^3x$) of finding the particle at one particular point, including the singularity, is always zero. So merely considering the quantum mechanical probability distribution is not helpful. However, what seems more relevant is whether the probability density flows into or out of the singularity, i.e., whether the probability flux into the singularity is zero or not. If there is a decreased change over time to find the particle outside the singularity, it seems natural to conclude that the particle might have ended up in the singularity. If there is no flux into or out of the classical singularity, then the singularity is avoided quantum mechanically. Concretely, in this context this condition means that{\footnote{One might worry that the integral in \eqref{2} is not finite because it involves an integration over an unbounded surface $C_R$. It might therefore be worth considering alternative criteria like demanding that \eqref{2} holds for appropriate segments of $C_R$ or that $\limsup_{R \to 0} \sup_{x \in C_R}  |n_\mu({\bf x}) j^\mu({\bf x})| =0$.}} 
\be
\lim_{R \to 0} \oint_{C_R} \ud\sigma({\bf x}) |n_\mu({\bf x}) j^\mu({\bf x})| =0 ,
\label{2}
\en 
where $C_R$ is the cylinder with radius $R$ and normal field $n_\mu$ centered around the line $r=0$ in space-time and $ j^\mu = {\bar \psi} \gamma^\mu \psi$ is the Dirac current. (The absolute value is taken so that there is no incoming and no outgoing flux.) For any state, the condition \eqref{2} is actually satisfied since the $L_2$-norm of the wave function is preserved. So in this case the notion ties together with the unitarity of the quantum dynamics.

When it comes to quantum gravity, there is no immediate probability distribution that could be used to formulate a Born rule, even in the case of minisuperspace models, due to the constraint nature of the theory. Nevertheless, there is a conserved current and we can consider whether there is flux into singular metrics. This conserved current can be derived as a Noether current corresponding to the $U(1)$ symmetry of the action $S(\psi) = \langle \psi | {\widehat H}\psi \rangle$ from which the Wheeler-DeWitt equation ${\widehat H}\psi =0$ is derived (provided there is a kinematical inner product with respect to which $ {\widehat H}$ is symmetric). As a no-singularity criterion we now consider the condition that there is no flux into singular metrics.

An advantage is that we do not need to consider a Hilbert space or deal with the problem of time. Another advantage is that the no-flux criterion can be naturally written in coordinate-free language using differential forms. Namely, in the case of an $n$-dimensional minisuperspace, the local conservation of the current $J$ yields that $J$ is most naturally thought of as an instance of a closed $(n-1)$-form. The flux of such a form through codimension-1 surfaces $\partial \Omega$ (encompassing the singular metrics) is intrinsically defined by the integral $\int_{\partial \Omega}J$.

On the other hand, the conserved current may not directly be related to probability flow. While in the case of the Dirac theory the no-singularity criterion was related to conservation of probability, this may no longer be so in the case of quantum gravity. The situation may be compared to that of the Klein-Gordon equation describing a single spinless particle. The conserved current $j^\mu \sim {\textrm{Im}}(\psi^* \pa^\mu \psi)$ is not a probability current, but rather is often interpreted as the charge current. Nevertheless the flux of this current can be considered even though it may not be sufficient to completely capture the notion of singularity avoidance. 

This being said, in the semi-classical regime, the current is sometimes treated as a probability current \cite{hawking86,vilenkin89,halliwell91b}. Also, our notion of singularity avoidance is relevant in the case of Bohmian gravity. In Bohmian quantum gravity \cite{pinto-neto18}, there is an actual metric and actual matter degrees of freedom, such as a scalar field, and the integral curves of the conserved current form the possible trajectories. So in this case, the classical notion of singularity can be employed and singularities are typically avoided when there is no flux through the singularities. The situation is similar in case of the Klein-Gordon equation, where the Bohmian trajectories correspond to the integral curves of the Klein-Gordon current. In addition, in the context of Bohmian non-relativistic quantum mechanics, the $|\psi|^2d^3x$-probability for a particle to run into the singularity of the Coulomb potential is zero if and only if the no-flux condition is obeyed \cite{berndl95c,teufel05}.   

In the next section, we will consider a particular minisuperspace model corresponding to a Friedmann-Lema\^itre-Robertson-Walker (FLRW) universe with dust, where the dust is described by the Brown and Kucha{\v r} method \cite{brown95}. In this example, it turns out that the chosen operator ordering plays a crucial role in whether or not singularities occur.

\section{minisuperspace with dust} \label{dust}
In the Brown-Kucha{\v r} description \cite{brown95,maeda15}, dust is described by a matter fluid determined by the rest mass density $\rho$ and a 4-velocity field $U^\mu$ which can be parametrized by certain (noncanonical) scalar fields. The classical action including gravity is{\footnote{We take $\hbar=c=1$.}}
\be
S =  \frac{1}{2} \int \ud^4x \sqrt{-g} \rho \left( g^{\mu\nu} U_\mu U_\nu - 1 \right) + \frac{1}{16\pi G} \int \ud^4x \sqrt{-g} R  ,
\en
where $G$ is the gravitational constant, $g_{\mu \nu}$ is the 4-metric and $R$ is the Ricci scalar. Assuming spatial homogeneity and isotropy, the metric is given by the FLRW metric
\be
\dd s^2 = N(t)^2 \dd t^2 - a(t)^2  \dd \Omega^2 ,
\label{5}
\en
where $N>0$ is the lapse function, $a$ the scale factor, and $\dd \Omega^2$ is the spatial line-element on 3-space. We assume that there is no spatial curvature and that 3-space is compact, with comoving volume $V$. Because of the symmetry, the dust field $U^\mu$ can be parametrized by a single scalar field $T=T(t)$ as $U_\mu = \pa_\mu T$. The effective classical Lagrangian is given by 
\be
L = V\left[  \frac{1}{2} N a^3 \rho \left( \frac{\dot T^2}{N^2} - 1\right)  - \frac{1}{2N\ka^2}a \dot a^2 \right],
\en
where $\kappa = \sqrt{4\pi G/3}$. The corresponding classical equations of motion are
\be
\dot T^2 = N^2 , \quad  \frac{\ud}{\ud t}\left(a^3 \rho \right)=0 , \quad \frac{\dot a^2}{N^2a^2} = 2\ka^2 \rho, \quad \frac{\ud}{\ud t}\left(\frac{{\dot a} a^2 }{N} \right)= \frac{3}{2} \frac{a{\dot a}^2}{N}.
\en
The lapse function is an arbitrary function of time, which relates to the time reparametrization invariance. In the gauge $N=1$, so that $t$ is cosmic time (i.e., the proper time for an observer moving with the expansion of the Universe), we can write these equations as
\be
 T = t + t_0, \quad  \frac{\ud}{\ud t}\left(a^3 \rho \right)=0 , \quad  \frac{\dot a^2}{a^2} = 2\ka^2 \rho, \quad \frac{\ddot a}{a} = - \ka^2 \rho .
\en
Without loss of generality, we can put the constant $t_0=0$, so that the matter field $T$ just equals cosmic time. The resulting equations are the familiar ones for dust, which leads to the following evolution of the scale factor:
\be
a(t) = (c_1 t + c_2)^{2/3} ,
\en
with $c_1,c_2$ constant. So, classically, there is always a big bang or big crunch singularity, when $a=0$, obtained at $t= -c_2/c_1$.

Canonical quantization of this classical theory yields the Wheeler-DeWitt equation
\be
{\widehat H}\psi:=\ii \pa_T \psi -  \frac{1}{2M} \frac {1}{ a^{m+1}}  \pa_a \left( a^m \pa_a \psi \right)  = 0,
\label{15}
\en
where $\psi=\psi(a,T)$, and hence $\psi$ is independent of time $t$, and $M=V/\ka^2$. The variable $m$ corresponds to a choice of operator ordering.\footnote{Of course, more general operator orderings could be considered, such as
\be
\ii\frac{1}{w(T)} \partial_{T}\left(w(T) \psi\right)+\frac{1}{2Maf(a)}\partial_{a}\left(f(a)\partial_a\psi\right) = 0, 
\en
which is still a local constraint.
} The choice $m=-1/2$ corresponds to using the Laplace-Beltrami operator corresponding to the DeWitt metric on minisuperspace \cite{maeda15}. Other operator orderings have been considered in, e.g., \cite{amemiya09}.

There is a conservation equation
\be
\pa_T  j_T + \pa_a j_a = 0,
\en
where
\be
j_T =  |\psi|^2 a^{m+1}  , \qquad j_a = -\frac{1}{2M} a^m{\text{Im}}(\psi^*\pa_a \psi) .
\en
Avoidance of the singularity now means that there is no flux through $a=0$ in the half-plane determined by $a\in{\mathbb R}^+:=(0,+\infty), T \in {\mathbb R}$}. 
In other words,{\footnote{In the Bohmian theory, the dynamics is given by 
\be
\dot T = N \frac{j_T}{|\psi|^2a^{m+1}} , \qquad \dot a = N \frac{j_a}{|\psi|^2a^{m+1}} ,
\en
with $N$ as an arbitrary function of time \cite{pinto-neto18}. This dynamics preserves the measure $|\psi(a,T)|^2a^{m+1}d adT$ (which is no probability measure). In this case, the no-flux condition \eqref{noflux} is closely related to the property that the $|\psi(a,T)|^2a^{m+1}d adT$-measure of the set of ``initial'' configurations that start or end up in $a=0$ is zero.}}
\be
\lim_{a\to 0}  \int_{-\infty}^{+\infty} \ud T\,|j_a (a,T)| = 0.
\label{noflux}
\en 

Using $x = \frac{2}{3}a^{3/2}$ and $\tau=-T$, the Wheeler-DeWitt equation \eqref{15} reduces to 
\be
\ii \pa_\tau \psi =  -\frac{1}{2M} \frac{1}{x^n}\pa_x x^n \pa_x \psi ,
\label{sch0}
\en
with $n= 2m/3+1/3$. Since $a\in{\mathbb R}^+$, we have that $x\in{\mathbb R}^+$. Using the transformation $\psi = x^{-n/2} \phi/(2/3)^n$, we get
\be
\ii \pa_\tau \phi =  -\frac{1}{2M} \pa^2_x \phi + \frac{c}{2M} \frac{1}{x^2} \phi,
\label{sch}
\en
with $c= (n/2 - 1) n/2$. So we get the non-relativistic Schr\"odinger equation with a $1/x^2$ potential. We have that
\be
- j_a  =  j_x = \frac{1}{2M} {\text{Im}}(\phi^*\pa_x \phi),
\en
with $j_x$ the usual probability current associated with the Schr\"odinger equation \eqref{sch}, which satisfies
\be
\pa_\tau |\phi(x,\tau)|^2 + \pa_x j_x (x,\tau) = 0.
\label{cont}
\en
The criterion that there be no flux into the big bang configurations $(a=0)$ now reads 
\be
\lim_{x\to 0}  \int_{-\infty}^{+\infty} \ud \tau\,|j_x (x,\tau)| = 0.\footnote{
	Similarly, a no-flux criterion might be formulated to avoid big rip singularities (which classically correspond to $a \to \infty$ in finite time).}
\label{noflux2}
\en

The form of the Schr\"odinger equation \eqref{sch} suggests that one look for a \textit{unitary} dynamics on the Hilbert space $L^2(\mathbb{R}^+)$, with $\tau$ playing the role of time \cite{maeda15,pal16}. A unitary dynamics will preserve the norm of $\phi$, i.e., 
\be
\|\phi(x,\tau) \|_2 = \left( \int^{+\infty}_0  |\phi(x,\tau)|^2\ud x \right)^{1/2}
\en
is constant as a function of $\tau$, i.e., $\pa_\tau \|\phi \|_2 = 0$. On the other hand, from the continuity equation \eqref{cont} we have 
\be
\pa_\tau \|\phi(x,\tau) \|^2_2 =  -\int^{+\infty}_0  \ud x \pa_x j_x (x,\tau) = - \lim_{x\to 0} j_x (x,\tau)
\label{110}
\en
(provided $\lim_{x\to +\infty}j_x (x,\tau) = 0$). Hence, $\lim_{x\to 0} j_x (x,\tau) =0$ and there is no flux in each finite $\tau$ interval. This does not necessarily imply the no-flux condition \eqref{noflux2}, since there might still be flux for $|\tau|$ going to infinity. However, we will not worry about this since classically already, $|\tau|$ to infinity corresponds to infinite proper cosmic time.{\footnote{In the Bohmian theory, $\lim_{x\to 0} j_x (x,\tau) =0$ is enough to ensure that trajectories do not reach the singularity in finite cosmic proper time.}}

This being said, we will refrain from demanding a unitary dynamics. After all, time does not appear in the theory and promoting the variable $\tau$ to play the temporal role relies on a certain way to solve the problem of time. At this stage we do not want to commit ourselves to any solution to that problem. So, we want to regard \eqref{sch} as a partial differential equation at face value, without regarding $\tau$ as a time variable. Requiring that solutions $\phi$ are square integrable over the $(x,\tau)$ half-plane ($ x>0$) is too strong, but in order to preserve the possibility of a notion of (conditional) probability, it remains natural to require that they are square integrable over certain sections of the half-plane. 

In the case $c=0$, a $L^2(\mathbb{R}^+)$-integrability condition for $\phi(.,\tau)$ does not yet accomplish ruling out flux through $a=x=0$: the dynamics in this case can be seen as the restriction to the half-line of the free Schr\"odinger evolution over the whole line. For a generic solution of the latter problem, there will be flux through the origin. Extra boundary conditions need to be imposed to prohibit such a flux. (Requiring a unitary dynamics on the half-line yields the boundary condition $\phi|_{x=0} = C \pa_x \phi|_{x=0}$, with $C \in {\mathbb{R}} \cup \{\infty\}$. In this case, we have a again $\lim_{x\to 0} j_x (x,\tau) =0$.) 

In the case $c>0$, the representation \eqref{sch} of the Schr\"odinger equation exhibits a repulsive potential term proportional to $1/x^2$. One might wonder whether this potential, which is in principle infinitely repelling at the origin, is sufficient to prohibit flux through $a=x=0$. This appears to be indeed the case, under mild assumptions, when $c\geq 3/4$. For example, if we require $\phi$ and ${\widehat h}\phi := -\frac{1}{2M} \pa^2_x \phi + \frac{c}{2M} \frac{1}{x^2} \phi$ to be in $L^2(\mathbb{R}^+)$,{\footnote{This condition guarantees that $\pa_x j_x = 2{\text{Re}} (\ii \psi^* {\widehat h} \psi)$ is integrable so that the argument concerning \eqref{110} can be made in the first place.}} then $\phi = \mathcal{O} (x^{3/2})$ and $\pa_x \phi = \mathcal{O} (x^{1/2})$ \cite[pp.\ 249-250]{gitman12}, so that $\lim_{x\to 0} j_x (x,\tau) =0$. (Clearly, for $c=0$, the condition that $\phi$ and ${\widehat h} \phi$ are in $L^2(\mathbb{R}^+)$ is not sufficient to guarantee no flux.) The condition on $\psi$ could be even be weakened, but this will be discussed in future work. 

The requirement that $c\geq 3/4$ relates to that fact that for those values are those for which ${\widehat h}$ is essentially self-adjoint \cite{reed75}. By virtue of Stone's theorem, such essential self-adjointness corresponds to a situation where a unique unitary dynamics $e^{-\ii\tau {\widehat h}}$ can be associated to the operator ${\widehat h}$. 

If $T$ is taken to be time, then one could take $J_T$ to be a probability distribution by requiring normalizability. Then the current corresponds to a probability current. This is also the case in the Bohmian theory, since it follows from the Bohmian dynamics that $T$ can be treated as a clock variable.

To conclude, by assuming a certain operator ordering together with a mild integrability condition can rule out any flux into singular configurations.

\section{Conclusion}\label{conclusion}
We investigated the role of operator orderings in the question of space-time singularities corresponding to a big bang or big crunch. We used the criterion that quantum mechanically there are space-time singularities if and only if there is quantum flux into singular 3-metrics. For a quantum minisuperspace model with dust, an operator ordering can be chosen in the quantization process such that big bang or big crunch singularities are avoided, without assuming any boundary conditions. This is because of the appearance of an effective potential that repels from the singularity. This potential falls off rapidly away from the singularity and hence merely seems to affect the quantum dynamics {\em near} the singularity. Effects {\em away} from the singularity, such as the appearance of an effective cosmological constant, described in \cite{demaerel19a}, are presumably unaffected by such alternative operator orderings. The situation is similar to that in loop quantum gravity where the quantum effects also cause a bouncing behavior near the singularity \cite{bojowald01,kiefer04,ashtekar06b,ashtekar06c,ashtekar08}.

Perhaps singularities can similarly be avoided in other minisuperspace models, e.g., with a canonical scalar field, or in full quantum gravity. In view of the absence of experimental evidence for one or the other operator ordering, the criterion that singularities are prohibited can perhaps narrow down the possibilities. 

In the example of minisuperspace with dust, our singularity criterion actually fits well with the usual way of dealing with the problem of time. But it remains to be seen if this is still the case for other minisuperspace models.

\section{Acknowledgments}
It is a pleasure to thank Wojciech De Roeck, Detlef D\"urr, and Christian Maes for valuable discussions. W.S.\ is supported by the Research Foundation Flanders (Fonds Wetenschappelijk Onderzoek, FWO), Grant No.\ G066918N.


\begin{thebibliography}{10}
\bibitem{christodoulakis86}
{T.\ Christodoulakis and J.\ Zanelli, ``Operator Ordering in Quantum Mechanics
  and Quantum Gravity'', {\em Nuovo Cimento B} {\bf 93}, 1-21 (1986).}

\bibitem{halliwell91b}
{J.J.\ Halliwell, ``Introductory lectures on quantum cosmology'', in {\em
  Quantum Cosmology and Baby Universes}, eds.\ S.\ Coleman, J.B.\ Hartle, T.\
  Piran and S.\ Weinberg, World Scientific, Singapore, 159-243 (1991) and
  arXiv:0909.2566 [gr-qc].}

\bibitem{kontoleon99}
{N.\ Kontoleon and D.L.\ Wiltshire, ``Operator ordering and consistency of the
  wavefunction of the Universe'', {\em Phys.\ Rev.\ D} {\bf 59}, 063513 (1999)
  and arXiv:gr-qc/9807075.}

\bibitem{steigl06}
{R.\ {\u S}teigl and F.\ Hinterleitner, ``Factor ordering in standard quantum
  cosmology'', {\em Class.\ Quantum Grav.}\ {\bf 23}, 3879-3893 (2006) and
  arXiv:gr-qc/0511149.}

\bibitem{haga17}
{J.\ Haga and R.L.\ Maitra, SIGMA {\bf 13}, 039 (2017) and arXiv:1612.05622
  [math-ph].}

\bibitem{dewitt67a}
{B.S.\ DeWitt, ``Quantum Theory of Gravity. I. The Canonical Theory'', {\em
  Phys.\ Rev.}\ {\bf 160}, 1113-1148, (1967).}

\bibitem{lund73}
{F.\ Lund, ``Canonical Quantization of Relativistic Balls of Dust'', {\em
  Phys.\ Rev.\ D} {\bf 8}, 3253-3259 (1973).}

\bibitem{gotay83}
{M.J.\ Gotay and J.\ Demaret, ``Quantum cosmological singularities'', {\em
  Phys.\ Rev.\ D} {\bf 28}, 2402-2413 (1983).}

\bibitem{bojowald01}
{M.\ Bojowald, ``Absence of a Singularity in Loop Quantum Cosmology'', {\em
  Phys.\ Rev.\ Lett.}\ {\bf 86}, 5227-5230 (2001) and arXiv:gr-qc/0102069.}

\bibitem{kiefer04}
{C.\ Kiefer, {\em Quantum Gravity}, International Series of Monographs on
  Physics {\bf 124}, Clarendon Press, Oxford (2004).}

\bibitem{ashtekar06b}
{A.\ Ashtekar, T.\ Pawlowski and P.\ Singh, ``Quantum nature of the big bang'',
  {\em Phys.\ Rev.\ Lett.}\ {\bf 96}, 141301 (2006) and arXiv:gr-qc/0602086.}

\bibitem{ashtekar06c}
{A.\ Ashtekar, T.\ Pawlowski and P.\ Singh, ``Quantum nature of the big bang:
  An analytical and numerical investigation'', {\em Phys.\ Rev.\ D} {\bf 73},
  124038 (2006) and arXiv:gr-qc/0604013.}

\bibitem{ashtekar08}
{A.\ Ashtekar, A.\ Corichi and P.\ Singh, ``Robustness of key features of loop
  quantum cosmology'', {\em Phys.\ Rev.\ D} {\bf 77}, 024046 (2008) and
  arXiv:0710.3565 [gr-qc].}

\bibitem{kiefer10}
{C.\ Kiefer, ``Can singularities be avoided in quantum cosmology?'', {\em Ann.\
  Phys.\ (Berlin)} {\bf 19}, 211-218 (2010).}

\bibitem{blyth75}
{W.f.\ Blyth and C.J.\ Isham, ``Quantization of a Friedmann universe filled
  with a scalar field'', {\em Phys.\ Rev.\ D} {\bf 11}, 768-778 (1975).}

\bibitem{hawking86}
{S.W.\ Hawking and D.N.\ Page, ``Operator ordering and the flatness of the
  universe'', {\em Nucl.\ Phys.}\ B {\bf 264}, 185-196 (1986).}

\bibitem{vilenkin89}
{A.\ Vilenkin, ``Interpretation of the wave function of the Universe'', {\em
  Phys.\ Rev.\ D} {\bf 39}, 1116-1122 (1989).}

\bibitem{pinto-neto18}
{N.\ Pinto-Neto and W.\ Struyve, ``Bohmian quantum gravity and cosmology'', in
  {\em Applied Bohmian Mechanics: From Nanoscale Systems to Cosmology}, 2nd
  edition, eds.\ X.\ Oriols and J.\ Mompart, 607-656 (2019).}

\bibitem{berndl95c}
{K.\ Berndl, D.\ D\"urr, S.\ Goldstein, G.\ Peruzzi and N.\ Zangh\`\i, {\em
  Commun.\ Math.\ Phys.}\ {\bf 173}, 647-673 (1995) and
  arXiv:quant-ph/9503013.}

\bibitem{teufel05}
{S.\ Teufel and R.\ Tumulka, ``Simple Proof for Global Existence of Bohmian
  Trajectories'', {\em Commun.\ Math.\ Phys.}\ {\bf 258}, 349-365 (2005) and
  math-ph/0406030.}

\bibitem{brown95}
{J.D.\ Brown and K.V.\ Kucha{\v{r}}, ``Dust as a standard of space and time in
  canonical quantum gravity'', {\em Phys.\ Rev.\ D} {\bf 51}, 5600 (1995) and
  arXiv:gr-qc/9409001.}

\bibitem{maeda15}
{H.\ Maeda, ``Unitary evolution of the quantum Universe with a Brown-–Kushar
  dust'', {\em Class.\ Quantum Grav.}\ {\bf 32}, 235023 (2015) and
  arXiv:1502.06954.}

\bibitem{amemiya09}
{F.\ Amemiya and T.\ Koike, ``Gauge-invariant construction of quantum
  cosmology'', {\em Phys.\ Rev.\ D} {\bf 80}, 103507 (2009) and arXiv:0910.4256
  [gr-qc].}

\bibitem{pal16}
{S.\ Pal and N.\ Banerjee, ``Anisotropic models are unitary: A rejuvenation of
  standard quantum cosmology'', {\em J.\ Math.\ Phys.}\ {\bf 57}, 122502 (2016)
  and arXiv:1601.00460 [gr-qc].}

\bibitem{gitman12}
{D.M.\ Gitman, I.V.\ Tyutin and B.L.\ Voronov, {\em Self-adjoint Extensions in
  Quantum Mechanics}, Progress in Mathematical PHysics {\bf 62}, Springer, New
  York (2012).}

\bibitem{reed75}
{M.\ Reed and B.\ Simon, {\em Methods of modern mathematical physics II: Fourer
  Analysis, Self-Adjointness}, Academic Press, San Diego (1975).}

\bibitem{demaerel19a}
{T.\ Demaerel, C.\ Maes and W.\ Struyve , ``Cosmic acceleration from quantum
  Friedmann equations'', arXiv:1901.09767 [gr-qc].}

\end{thebibliography}
\end{document}